\documentclass[a4paper,11pt]{article}

\usepackage{pos}
\usepackage[outercaption]{sidecap}
\usepackage{caption}
\usepackage{xcolor,lineno}
\usepackage{makecell}
\usepackage{empheq}

\title{Search for cosmic rays in GRANDProto300}

\ShortTitle{Search for cosmic rays in GRANDProto300}

\author*[a,b]{Jolan Lavoisier}
\author[c,d]{Xishui Tian}
\author[a,e]{Kumiko Kotera}
\author[f]{Takashi Sako}
\author[g]{Hanrui Wang}
\author[h]{Mauricio Bustamante}
\onbehalf{for the GRAND Collaboration{\normalsize \\ \normalfont(a complete list of authors can be found at the end of the proceedings)}\\}

\affiliation[a]{Institut d'Astrophysique de Paris, CNRS UMR 7095, Sorbonne Universite, 98 bis bd Arago 75014, Paris, France}
\affiliation[b]{ILANCE, CNRS – University of Tokyo International Research Laboratory, Kashiwa, Chiba 277-8582, Japan}
\affiliation[c]{Department of Astronomy, School of Physics, Peking University, Beijing 100871, China}
\affiliation[d]{Sorbonne Université, Université Paris Diderot, Sorbonne Paris Cité, CNRS, Laboratoire de Physique 5 Nucléaire et de Hautes Energies (LPNHE), 6 4 place Jussieu, F-75252, Paris Cedex 5, France} 
\affiliation[e]{IIHE/ELEM, Vrije Universiteit Brussel, Pleinlaan 2, 1050 Brussels, Belgium}
\affiliation[f]{Institute for Cosmic Ray Research, University of Tokyo, 5 Chome-1-5 Kashiwanoha, Kashiwa, Chiba 277-8582, Japan}
\affiliation[g]{National Key Laboratory of Radar Detection and Sensing, School of Electronic Engineering, Xidian University, Xi’an 710071, China}
\affiliation[h]{Niels Bohr International Academy, Niels Bohr Institute, University of Copenhagen, 2100 Copenhagen, Denmark}

\emailAdd{jolan.lavoisier@iap.fr}

\abstract{

GRANDProto300 (GP300) is a prototype array of the GRAND experiment, designed to validate the technique of autonomous radio-detection of astroparticles by detecting cosmic rays with energies between 10$^{17}$-10$^{18.5}$ eV. This observation will further enable the study of the Galactic-to-extragalactic source transition region. Between November 2024 and May 2025, 46 out of 300 antennas have been operational and collecting data stably. We present here our cosmic-ray search pipeline, which involves several filtering steps: (1) coincidence search for signals triggering multiple antennas within a time window, (2) directional reconstruction of events, (3) exclusion of clustered (in time and space) noise events, (4) polarization cut, (5) selection based on the size of the footprint, and (6) other less mature cuts in this preliminary stage, including visual cuts. The efficiency of the pipeline is evaluated and applied to the first batch of data, yielding a set of cosmic-ray candidate events, which we present.

\vspace{4mm}

}

\ConferenceLogo{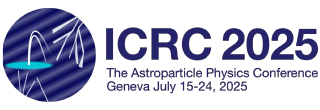}

\FullConference{39th International Cosmic Ray Conference (ICRC2025)\\  15--24 July, 2025\\ Geneva, Switzerland}

\begin{document}

\maketitle

\vspace{-0.3cm}
\section{Introduction}\label{sec1}
\vspace{-0.2cm}

The Giant Radio Array for Neutrino Detection (GRAND) experiment aims to detect cosmic particles with energies exceeding $10^{17}$ eV with an array of radio antennas~\cite{Martineau:GRAND_ICRC25}. GRANDProto300 (GP300) is a prototype array of the GRAND experiment, designed to validate the technique of autonomous radio-detection of astroparticles~\cite{GRANDprotoHW_25,Zhang-Kewen:electric_recons_ICRC25, Kato:exposure_ICRC25}. GP300 is specifically aimed at detecting cosmic rays with energies between 10$^{17}$-10$^{18.5}$ eV. This energy range is crucial for studying the transition region between Galactic and extragalactic sources of cosmic rays.

Within the GRAND collaboration, a data analysis pipeline is being developed to identify cosmic rays and analyze their properties, among the dataset of the first stable GP300 prototype runs (in the GRANDlib software framework~\cite{GRANlib24}). This proceeding focuses on the cosmic-ray identification procedure, which incorporates systematic discrimination based on air-shower signal and background noise properties: time and space clustering, polarization, size and shape of footprint, arrival directions, and additional manual cuts.

\vspace{-0.3cm}
\section{Experimental setup and dataset}\label{sec2}
\vspace{-0.2cm}

GP300 consists of 300 antennas spread over an area of 200\,km$^2$ situated in the Gobi desert, near Dunhuang, China (lying at $40.99434^\circ$N, $93.94177^\circ$E). As of July 2025, 65 antennas with an infill step of 577 m have been deployed and are operational, collecting data stably \cite{GRANDprotoHW_25}. Each Detection Unit (DU) operates with a sampling frequency of 500\,MHz and an analog-to-digital (ADC) converter digitizes signals across three distinct channels, each constructed from a dipole arm of the antenna, along a direction ($X$ channel along North-South, $Y$ channel along West-East, and $Z$ channel along the vertical axis). The ADC data, recorded over a duration of 2\,$\mu$s in a frequency range of 50 to 200\,MHz, is then converted to voltage units. The GRAND experiment handles data in three formats: Monitoring Data (MD), continuous monitoring of the background radio environments to identify and filter out noise; Unit Data (UD), which records signals from individual DUs that pass a local trigger at DU level based on 6 parameters, including two amplitude thresholds set at 5 and 3 sigma above background noise, ensuring that only transient radio pulses are captured \cite{Correa:NUTRIG_ICRC25, BenoitLevy:denoising_ICRC25}; Coincidence Data (CD), data from multiple ($\geq$4) antennas triggered within a 10\,$\mu$s time window, providing a more accurate identification and reconstruction of the signal. The collected data is processed via the GRANDlib software \cite{GRANlib24} in an offline treatment.

\begin{SCfigure}
    \vspace{-0.3cm}
    \centering
    \includegraphics[width=0.6\linewidth]{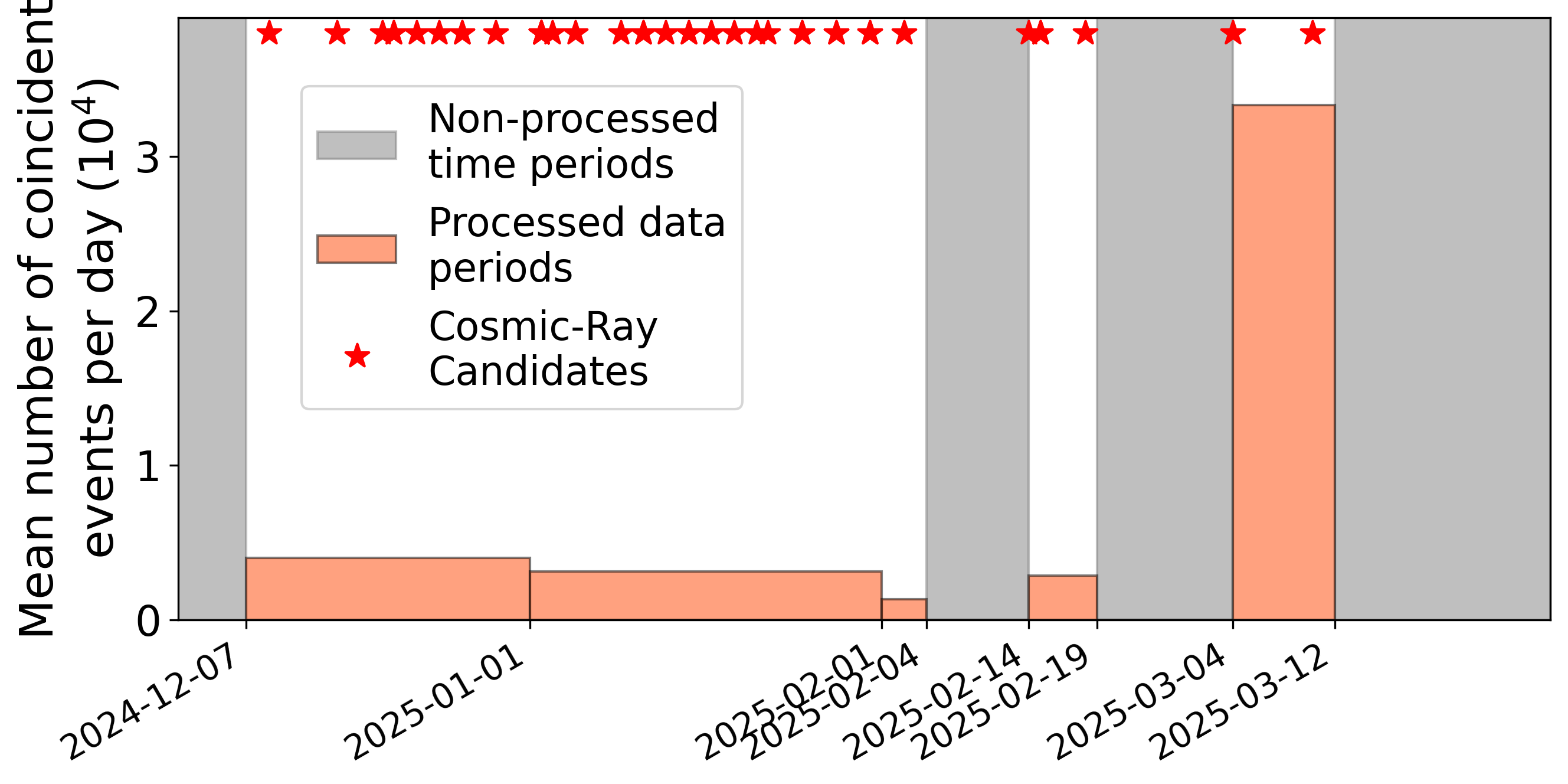}
    \caption{\footnotesize Mean number of coincident events per day (in units of $10^4$) in each data-processed periods (orange blocks) from Dec. 2024 to Mar. 2025. Gray areas indicate non-processed periods and red stars mark cosmic-ray candidate events selected via the pipeline described in this proceeding.}
    \label{fig:nb_event}
\end{SCfigure}

The data periods utilized for the search of cosmic rays are illustrated in Fig.~\ref{fig:nb_event}. The data set encompasses intervals during which the number of operational antennas and their trigger rate remained stable. Specifically, these periods include Dec. 7, 2024 to Feb. 4, 2025, Feb. 14 to 19, and Mar. 4 to 12. Throughout these intervals, the layout of the first deployed 46 antennas remained consistent. Note that the local trigger parameters were adjusted in March, resulting in an increased rate of events per day. In total, our dataset comprises $533,466$ coincident events ({\it i.e.} CD event).

\vspace{-0.2cm}
\subsection{Data processing: cosmic-ray search pipeline}
\vspace{-0.1cm}

The pipeline for cosmic-ray search is directly applied to CD events. They are subject to two types of noise: transient noise and stationary background noise. Transient noise induces events and is targeted for batch removal due to its high occurrence; for example, 70\,\% of all events in the period originate from an electric transformer in the northeast. Stationary background noise, on the other hand, appears within the baseline of the trace but does not independently trigger DUs. For this pipeline, we aim to adopt a conservative approach, prioritizing purity over efficiency in our cosmic-ray identification process.

\vspace{-0.2cm}
\subsection{Clustering Cut}\label{sec:cluster}

The GP300 dataset is affected by interference from commercial aircraft hovering near the array and regular emissions from an electric transformer located northeast of the site. This interference results in bursts of noise events that can be excluded from the analysis. Given that cosmic rays produce short radio pulses ($\lesssim 100$\,ns) and thus cannot result in multiple CD events in our data, a clustering algorithm is well-suited for this task.

\begin{figure}[t]
    \vspace{-0.3cm}
    \centering
    \includegraphics[width =0.48\textwidth]{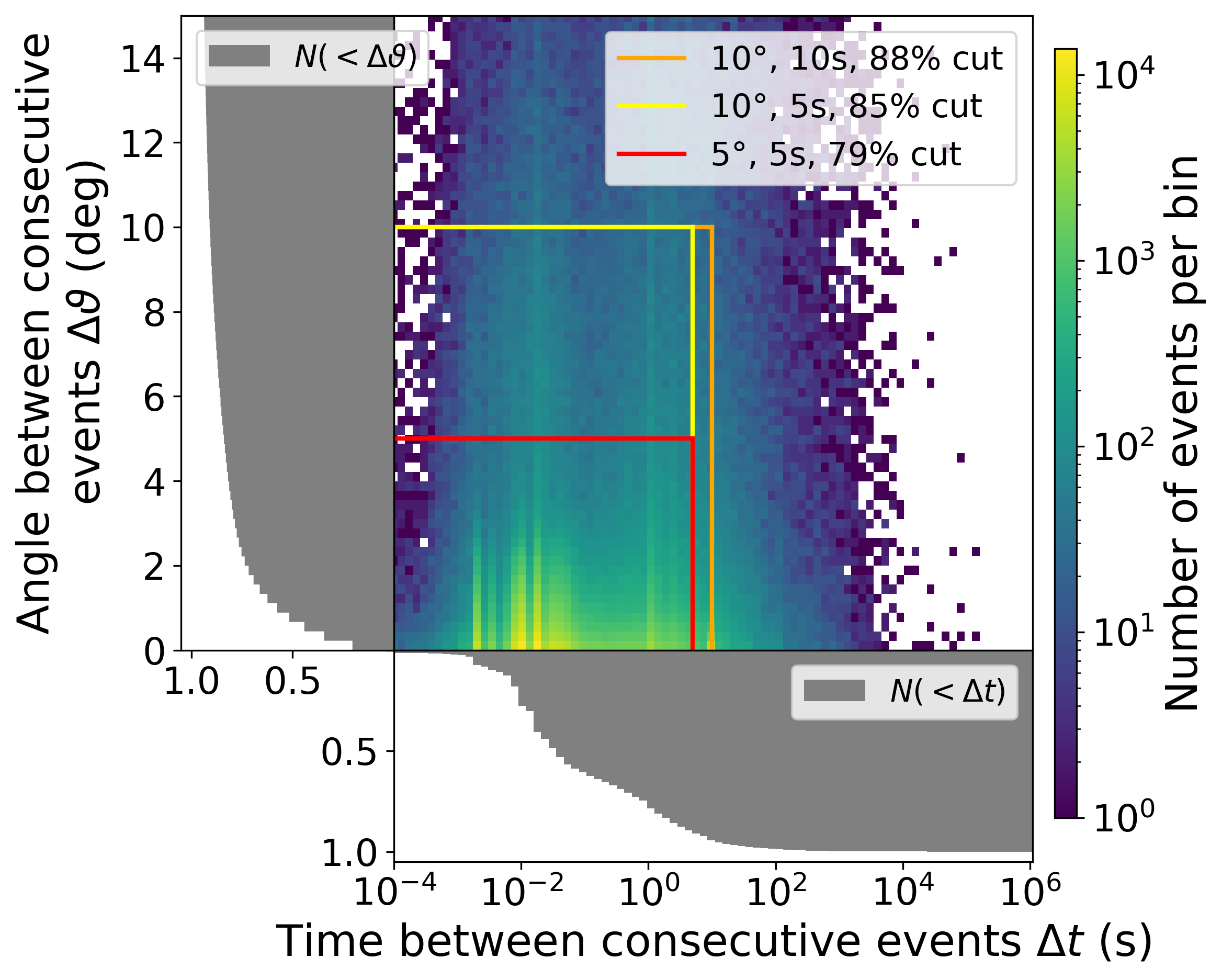}
    \includegraphics[width=0.48\textwidth]{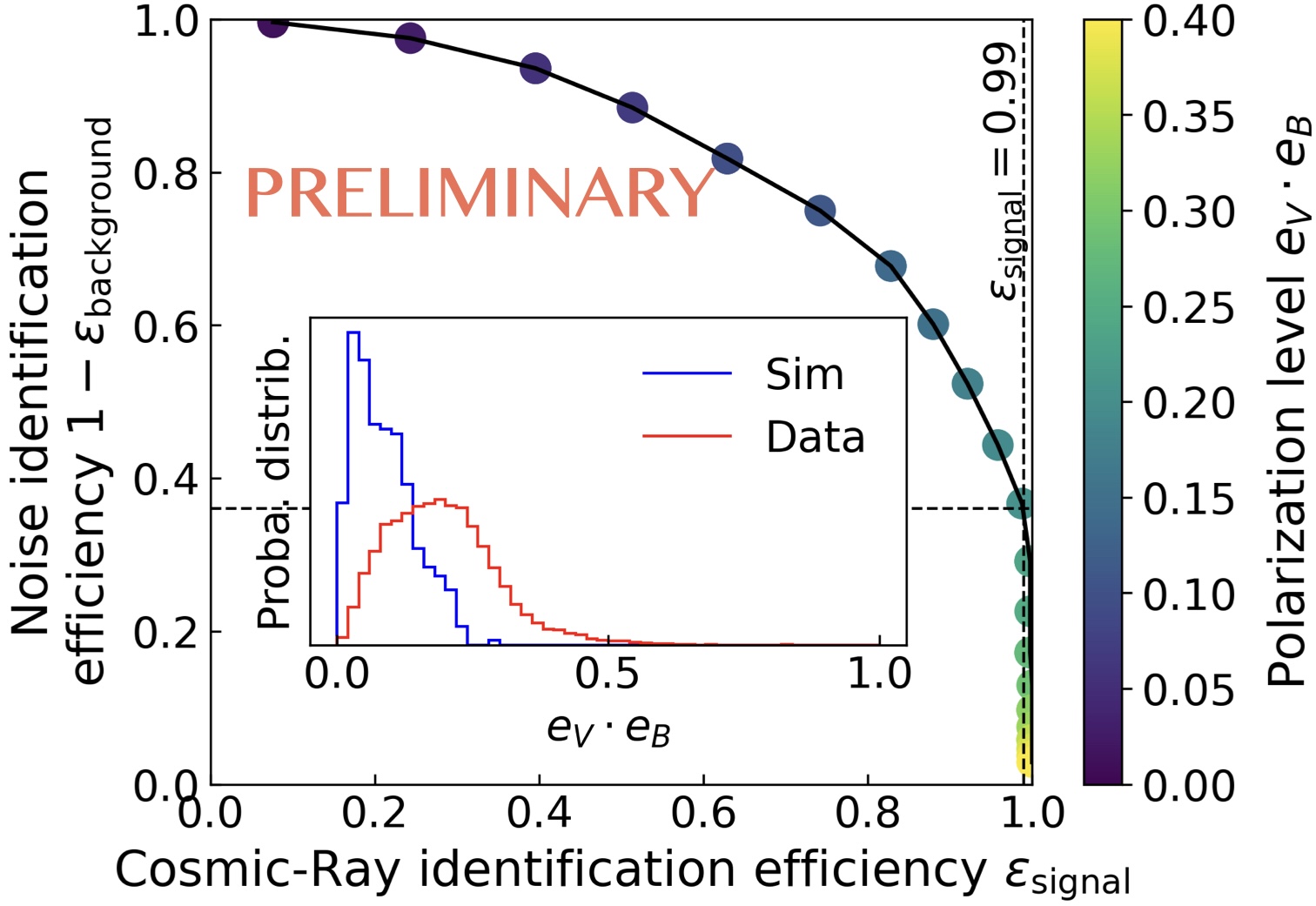}
    \caption{\footnotesize {\bf Left:} Clustering cut efficiency. Distribution of angular $\Delta \vartheta$ and temporal $\Delta t$ distances to the nearest neighbor events. The cumulative sums of these distributions are shown in gray. The red [yellow, orange] line indicates the regions in the parameter-space that corresponds to (and hence can be excluded) specific cut parameters sets ($\Delta \vartheta_{\rm cut}=5^\circ [10^\circ, 10^\circ]$ $\Delta t_{\rm cut}=5\,$s [5\,s, 10\,s]).
    {\bf Right:} Polarization cut efficiency. Fraction of background noise excluded by the cut versus fraction of signal included in the cut, depending on the tested cut parameter $\max (\bf{e}_V \cdot \bf{e}_B)$ (colorbar). The inset shows the probability distribution of the polarization level for simulated cosmic-ray signals (blue) and experimental data (noise in majority) (red). The vertical dashed line represents the chosen polarization cut threshold for cosmic-ray identification efficiency $\varepsilon_{\rm signal}=0.99$.}
    \label{fig:effvspur}
    \vspace{-0.5cm}
\end{figure}

The algorithm is based on the spatial proximity of events using Planar Wave Front (PWF) arrival direction reconstruction, following the method developed in ~\cite{Ferri_re_2025}. Multiple events occurring within the same angular and time window are designated as a cluster and consequently excluded from the dataset. The angular distance $\Delta \vartheta$ between events $A$ and $B$ is defined as the minimum between $\Delta\theta=|\theta_B - \theta_A|$ and $\Delta\varphi =|([\varphi_B-\varphi_A+\pi]\,\mod\,2\pi) -\pi|$. The temporal distance between $A$ and $B$ is defined as $\Delta t= |t_B-t_A|$. Fig.~\ref{fig:effvspur} (left) presents the distribution of angular and time distances for consecutive events in our dataset. By varying the angular and time cut parameters $(\Delta \vartheta_{\rm cut}, \Delta t_{\rm cut})$ below which the events are suppressed, the figure demonstrates the extent of data reduction achieved with different parameter choices. In this study, a conservative approach is adopted, and the parameters of $\Delta \vartheta_{\rm cut} = \min(\Delta \theta, \Delta \phi)=5^\circ$ and $\Delta t_{\rm cut}=5\,$s are retained, resulting in a cut of 79\,\% of the CD data, discriminated against as background noise, as seen in Fig.~\ref{fig:CRC} (right).

\vspace{-0.2cm}
\subsection{Polarization Cut}\label{sec:polar}
\vspace{-0.2cm}
The polarization cut offers the orthogonal approach of a cosmic-ray signal positive identification.
When a cosmic-ray air shower is created in the atmosphere, the geomagnetic field ${\bf B}$, through the Lorentz force, will act on the charged secondary particles and create a linear electric field along $\bf{k}\times\bf{B}$ where $\bf{k}$ is the particle velocity: this is the geomagnetic effect, the main source of radio emission from showers in air. This specific polarization signature along $\bf{k}\times\bf{B}$ can be used to identify cosmic-ray signals. In particular, \cite{Chiche_2022} proposed to use the component of the electric field along the {\bf B} direction ${\bf e}_V\cdot{\bf e}_B$, referred to as the {\it b}-ratio, as an estimator of polarization, where $\bf{e}_V$ is the normalized voltage vector from the antenna, $\bf{e}_B$ is the normalized geomagnetic field vector. ${\bf e}_V\cdot{\bf e}_B =0$ indicates that the electric field is orthogonal to $\bf{B}$, which is the case for cosmic rays, and ${\bf e}_V\cdot{\bf e}_B=1$ that the electric field is parallel to $\bf{B}$. This method demonstrates powerful identification efficiency: with a 5$\sigma$ trigger threshold on simulated voltage traces against Gaussian noise, 90\% of the antennas have a {\it b}-ratio $\lesssim 28$\%. This allows for the rejection of 72\% of noise events, assuming that noise events are uniformly distributed in the {\it b}-ratio.
Here we apply this method to our dataset. Given the challenges in accurately measuring polarity with DUs in this preliminary stage of the prototypes, we artificially fold the distribution to achieve a {\it b}-ratio ${\bf e}_V\cdot{\bf e}_B \lesssim 0.5$ in accordance with~\cite{Chiche_2022}, by choosing for the $\bf{e_V}$ vector an orientation towards positive $X$, $Y$ and $Z$. By doing so, we mitigate the risk of overlooking poorly reconstructed, polarized cosmic rays.
To apply this cut to an event, the estimator ${\bf e}_V\cdot{\bf e}_B$ is measured at each triggered antenna. The median value of each ${\bf e}_V\cdot{\bf e}_B$ over the set of antennas is assigned to the event. Given that cosmic rays are expected to exhibit low ${\bf e}_V\cdot{\bf e}_B$ values (low electric field along {\bf B}), events with median values higher than a specified polarization cut parameter $({\bf e}_V\cdot{\bf e}_B)_{\rm cut}$ are excluded. The optimization of this cut parameter is discussed in the following.

The efficiency of the polarization cut is evaluated using simulations as signals and data as noise. Simulations are run using ZHAireS to model 1000 extended air showers for half proton, half iron primaries, with energies ranging from 10$^{16.5}$ to 10$^{18.6}$\,eV. The electric fields generated from these simulations are processed through the radio-frequency (RF) chain using GRANDlib~\cite{GRANlib24}, and data-based stationary background noise is superimposed onto the obtained traces. 407 cosmic-ray simulations pass the local acquisition trigger settings. 35,167 collected data events are utilized as dummy noise for the analysis. To maximize the number of noise events in the data sample, only CD events originating from the directions of the plane and transformer are considered. The cumulative sum of simulations and the inverse cumulative sum of data are utilized to determine the optimal cut parameter. The polarization distribution obtained for cosmic rays and background noise are shown in the inset of Fig.~\ref{fig:effvspur} (right).

The polarization cut parameter $({\bf e}_V\cdot{\bf e}_B)_{\rm cut}$ is chosen to maximize cosmic-ray identification efficiency $\varepsilon_{\rm signal}$, {\it i.e.}\,identifying a cosmic ray as such (real positive), over noise identification efficiency $1 -\varepsilon_{\rm background}$ (real negative). Fig.~\ref{fig:effvspur} (right) shows that a polarization cut of $({\bf e}_V\cdot{\bf e}_B)_{\rm cut} =0.25$ provides a high signal efficiency $\varepsilon_{\rm signal}$ ($>99\,\%$) and discriminates up to 38\,\% of background noise events. The non-uniformity of the distribution of the {\it b}-ratio for background events, as illustrated in Fig.~\ref{fig:effvspur} (right), degrades the noise identification efficiency attained in more ideal simulations in \cite{Chiche_2022}. Applying this cut with the selected parameter yields the result in Fig.~\ref{fig:CRC} (right).

\vspace{-0.2cm}
\subsection{Systematic quality cuts}
\vspace{-0.2cm}

To ensure high-quality events, several additional quality cuts are implemented in the cosmic-ray search pipeline. They are independently applied to the data, and the order of the processes does not matter. We detail below all the additional cuts performed on the data. 
The performances of each cut, in terms of percentage of background events removed are presented in the right-hand table in Fig.~\ref{fig:CRC}. The cuts based on the antenna number (footprint size) and zenith angle (inclination of the shower) are built on physical grounds: GP300 focuses on specific cosmic-ray parameter ranges, that favor specific ranges in footprint size and zenith angle. The PWF error cut is more artificial as poorly reconstructed events could still be cosmic rays. However, in the current stage, events that cannot be reconstructed consistently with this PWF procedure are not usable (an error based on spherical wavefront is being implemented and tested). The last two cuts (SNR and RMS) are implemented at this stage of data commissioning in order to simplify  noise discrimination and help clean events emerge from the sample. They are however to be removed and/or refined for the next stage of the pipeline development, to process the next batch of data with the 65 operational antennas. 

\noindent {\bf Number of antennas.} This cut involves selecting events based on the size of the footprint, which refers to the area over which the radio signal is detected. This footprint size provides valuable information regarding the energy and type of the cosmic ray. We thus exclude all events with fewer than $N_{\rm DU, min} = 5$, as shown in Fig.~\ref{fig:CRC} (right).

\noindent {\bf Zenith.} Given its antenna sparsity, GP300 specializes in detecting highly inclined air showers with zenith angle $\theta \gtrsim 60^\circ$. The final cut removes events with a reconstructed arrival direction $\theta\le 60^\circ$ in zenith angle. In addition, reconstruction of near-horizon event is difficult, hence for now we remove as well events with zenith angle $\theta\ge 88^\circ$. Results are shown in Fig.~\ref{fig:CRC} (right).

\noindent {\bf PWF error.} 
This cut is designed to eliminate events reconstructed from multiple sources. Such events typically exhibit substantial angular errors. However, caution is required as this error is based on the PWF approximation. It does not compute an error on the arrival direction but rather on the wavefront shape, where a more planar wave shape results in a smaller error. The events with reconstructed angular error higher than ${\rm err_{PWF, max}}=0.5^\circ$ are excluded (Fig.~\ref{fig:CRC}, right).

\noindent {\bf Signal-to-Noise Ratio (SNR).}  We arbitrarily focus on events that exhibit high levels of SNR in their traces, especially along the $Y$-axis (West-East), SNR$_Y$, due to the direction of the geomagnetic field~\cite{Charrier_2019}. This cut enables us to identify the cleanest cosmic-ray events, for which reconstruction will also be more straightforward. We limit our sample to events where all DUs have SNR$_Y \ge 5$. The result of such cut is shown in Fig.~\ref{fig:CRC} (right). 

\noindent {\bf RMS.}  An additional quality cut involves limiting high Root Mean Squared (RMS) values of the $X$ and $Y$ 
channels for specific frequency bands. Given that the electric transformer exhibits significant emissions in identified frequency bands, this cut is designed to eliminate residual noise events stemming from this source. The proximity of the transformer to the east induces a stronger signal in the $X$ channel. The RMS limits are set by combining 3 conditions: $[{\rm RMS}_{X,Y}(50-80\,{\rm MHz}) < 2\,{\rm mV} ] \,\&\& \,[{\rm RMS}_{ X}(160-225\,{\rm MHz}) < 1\,{\rm mV} ]\, \&\& \, [{\rm RMS}_Y\,(160-225\,{\rm MHz}) < 3\,{\rm mV}]$.

\subsection{Manual quality cuts}
At this preliminary stage of the prototype and data analysis, in addition to the systematic cuts, manual (mostly visual) cuts are employed to further refine the search for cosmic-ray candidates, once the systematic search pipeline described above has been applied to the data.

 \noindent {\bf Trace visual cut.} This visual cut is based on the shape of the signal trace in each antenna. Cosmic rays are characterized by short-duration pulses (between 50 and 100\,ns) and low-frequency signals. 
 
 \noindent {\bf Footprint visual cut.} The second visual cut utilizes the spread and shape of the footprint, which refers to the area over which the radio signal is detected. The size of this footprint provides valuable information regarding the energy and type of the cosmic ray. Specifically, the triggered antennas should form a concentrated footprint of 5 to 10 antennas (see ~\cite{a2024pruningtooloptimizelayout}).

\noindent {\bf Timing cut.} A third cut involves comparing the measured arrival times with the expected times applying the PWF method~\cite{Ferri_re_2025}, assuming that the event is a cosmic ray. This ensures that no events constructed from signals coming from multiple sources are selected.

\begin{figure}[t]
    \vspace{-0.3cm}
    \centering
    \includegraphics[width =0.7\textwidth]{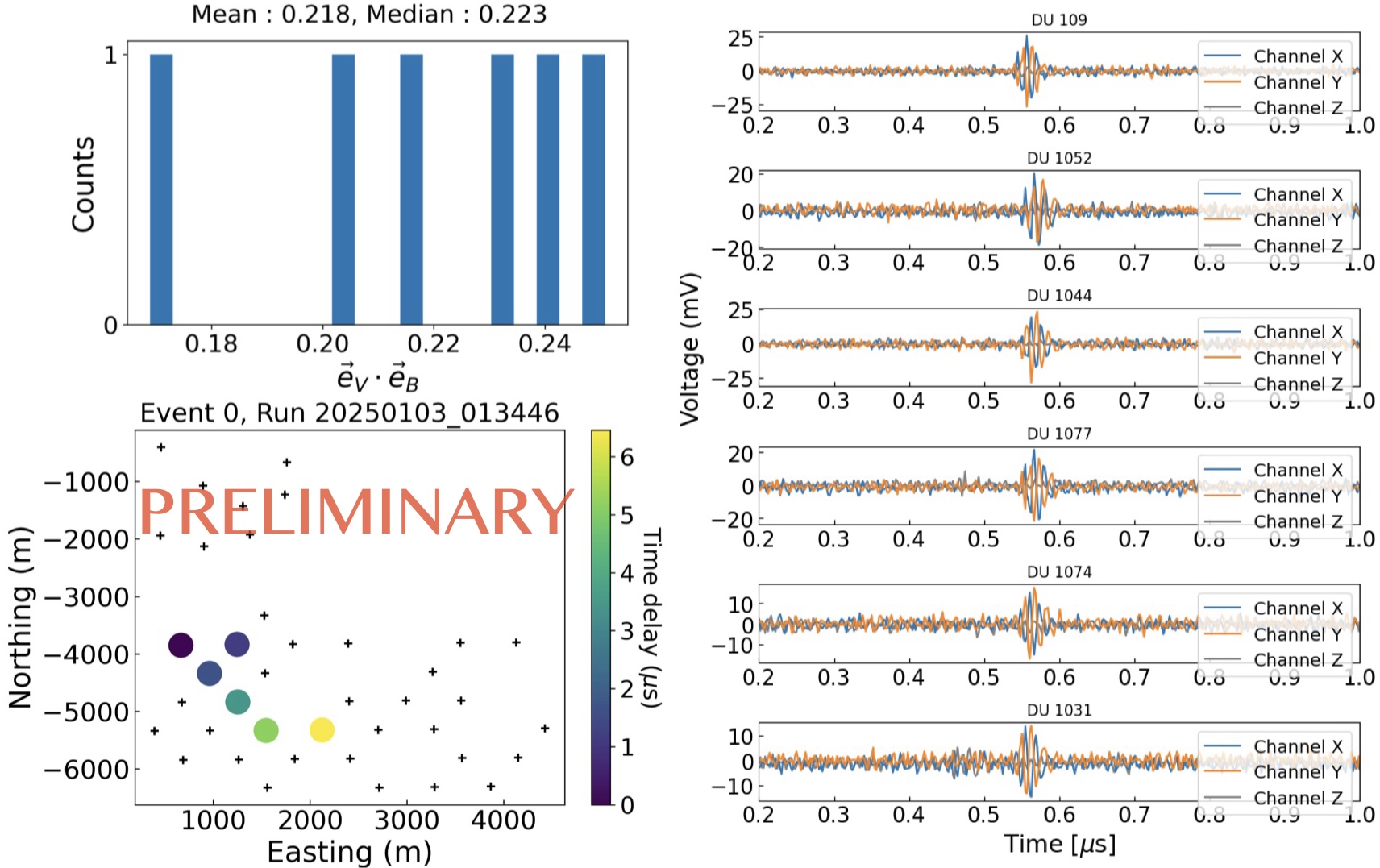}
    \caption{\footnotesize Examples of event visualization, for a cosmic-ray candidate that passed the entire pipeline, including manual cuts. The displayed information includes for each panel (top left) the distribution of polarization levels $({\bf e}_{V}\cdot {\bf e}_B)$ of the signal at triggered antennas in the event (mean and median values for the event indicated at the top), (bottom left) the position of the triggered antennas (antenna positions are indicated by crosses, and with triggered DUs in color, corresponding to the delay in arrival time as respect to the first triggered DU), and (right) the associated signal traces measured at each triggered antenna, for the 3 channels, over an acquisition time of $2\,\mu$s. }
    \label{fig:visual_good}
    \vspace{-0.5cm}
\end{figure}

\begin{SCfigure}
    \centering    
    \includegraphics[width=0.6\textwidth]{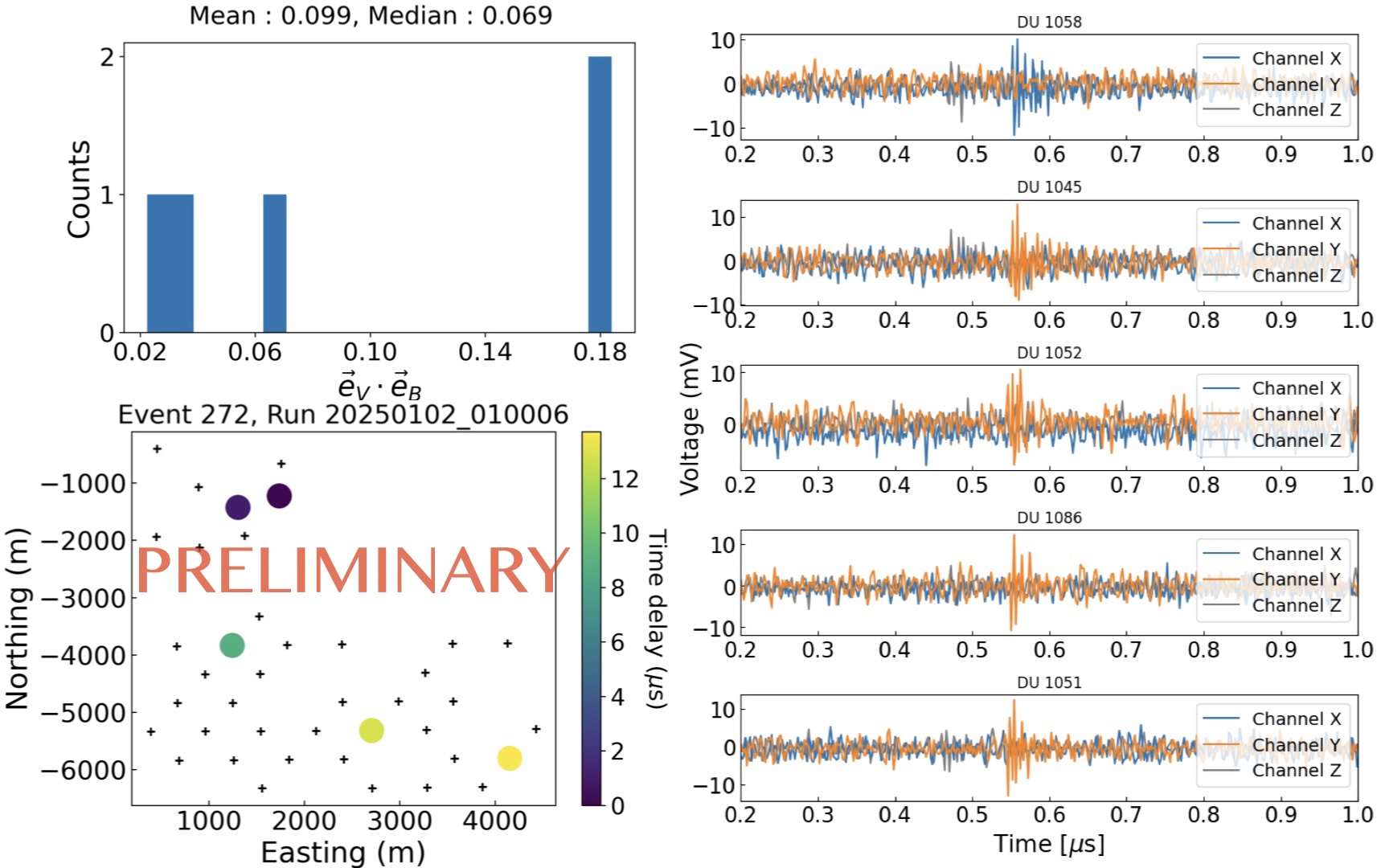}
    \caption{\footnotesize Same as Fig.~\ref{fig:visual_good}, but example of a noise event that passed earlier systematic cuts but do not pass the visual cut, due to non-compact footprint and trace shapes.}
    \label{fig:visual_bad}
        \vspace{-0.3cm}
\end{SCfigure}

Figs.~\ref{fig:visual_good} and \ref{fig:visual_bad} showcase two examples of events that passed the previous systematic cuts. The first example (Fig.~\ref{fig:visual_good}) represents a cosmic-ray candidate. This candidate is characterized by an elliptical footprint encompassing six DUs and a clean signal with short-duration pulses and low frequencies. In contrast, the second example (Fig.~\ref{fig:visual_bad}) depicts an event excluded by visual cuts. This event exhibits a dispersed footprint and a high-frequency signal.

After passing all cuts, the candidates are reconstructed using three different methods: lateral distribution function ~\cite{Gulzow:ICRC25} from deconvolved electric field reconstruction ~\cite{Zhang-Kewen:electric_recons_ICRC25}, angular distribution function (ADF) directly from voltage traces and electric field~\cite{Guelfand:volt_recons_ICRC25, guelfand2025reconstructioninclinedextensiveair}, and Graph Neural Networks directly from voltage traces ~\cite{Ferriere:GNN_ICRC25}. The majority of the reconstructions confirm the detection of cosmic rays in the expected energy range. These reconstruction methods serve as a basis for an additional quality cut, using multiple criteria: $\geq 5$ DUs with reconstructed electric fields (~\cite{Gulzow:ICRC25} and ~\cite{Guelfand:volt_recons_ICRC25}), a reconstructed energy $<10^{20}$\,eV, and a chi-squared value associated with the ADF method lower than 25, as described in~\cite{Guelfand:volt_recons_ICRC25}.
\vspace{-0.3cm}

\section{First batch of cosmic-ray candidates}
\vspace{-0.3cm}

\begin{figure}[!ht]
\vspace{-0.7cm}
    \centering
    \noindent\begin{minipage}{0.42\textwidth}
    \includegraphics[width=\textwidth]{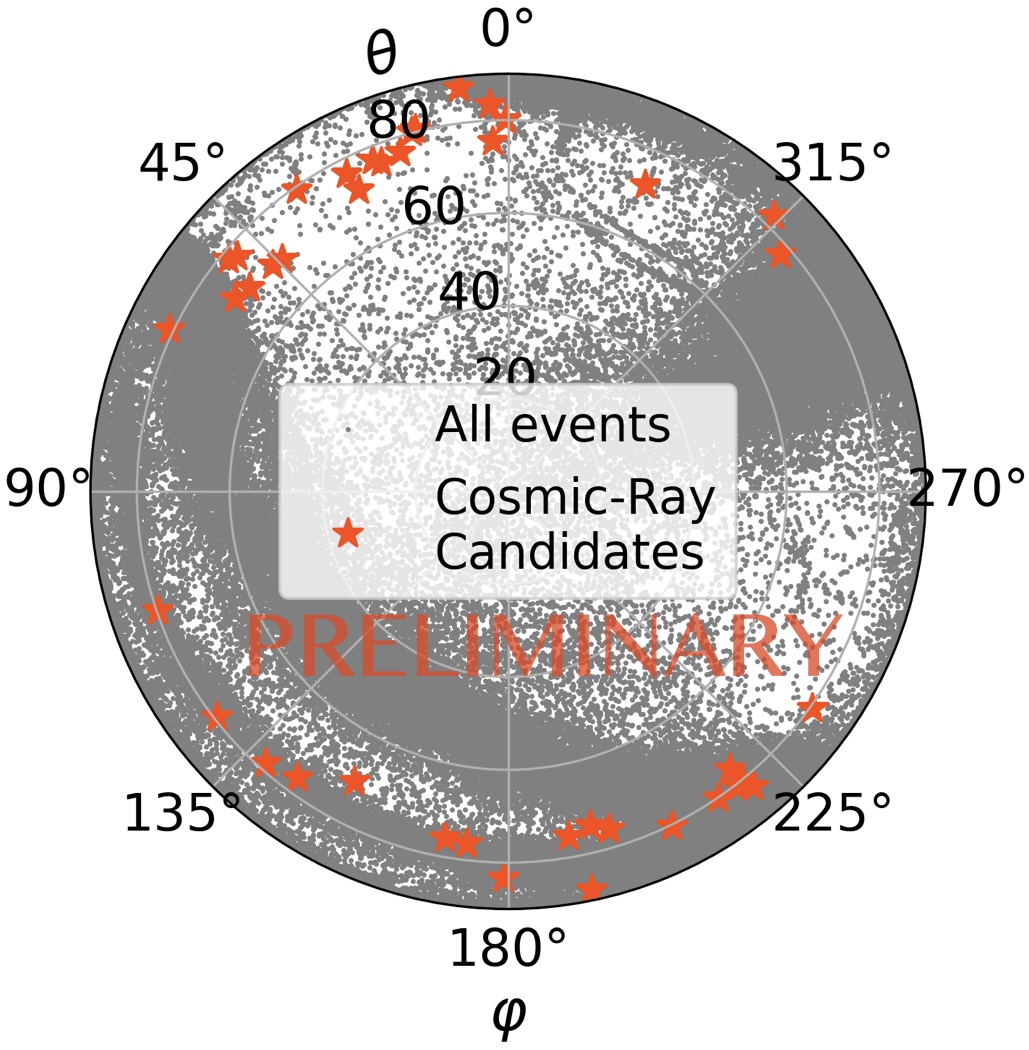}
    \end{minipage}
    \noindent\begin{minipage}{0.49\textwidth}
    \resizebox{\textwidth}{!}{%
        \begin{tabular}{lcr}\hline
            & Cut Parameters & Cut efficiency \\ \hline
            Clustering & \makecell{$\Delta t \leq 5\,$s \\ $\Delta \vartheta \leq 5^\circ$} & 79\,\% \\
            Polarization & $\mathbf{e_V \cdot e_B} < 0.25$ & 22\,\% \\
            Nb of DUs & $N_{\mathrm{DU}} \geq 5$ & 50\,\% \\
            Zenith & $\theta \in [60^\circ, 88^\circ]$ & 56\,\% \\
            PWF error & $\mathrm{err}_{\mathrm{PWF}} < 0.5^\circ$ & 31\,\% \\
            SNR & $\mathrm{SNR}_{\mathrm{Y}} \geq 5$ & 58\,\% \\
            RMS & \makecell{$\mathrm{RMS}_{X,Y}(50-80\,\mathrm{MHz}) < 2\,\mathrm{mV}$ \\ $\mathrm{RMS}_{X}(160-225\,\mathrm{MHz}) < 1\,\mathrm{mV}$ \\ $\mathrm{RMS}_{Y}(160-225\,\mathrm{MHz}) < 3\,\mathrm{mV}$} & 85\,\% \\ \hline
            Full pipeline & & $7 \times 10^{-5}$\,\% \\ \hline
        \end{tabular}}
    \end{minipage}
    \vspace{-0.3cm}
    
    \caption{\footnotesize {\bf Left:} Arrival directions (zenith $\theta$ and azimuth $\phi$ angles where $\theta=0$ indicates up and $\phi=0$ indicates north) of the cosmic-ray candidates in polar representation. {\bf Right:} Cut efficiencies of each process of our cosmic-ray search pipeline, applied to our dataset. The percentages indicate the proportion of events from the dataset that is excluded by each cut. The multiple methods are applied concurrently with an "and" logic, meaning that all criteria must be satisfied simultaneously to ensure robust validation. Hence the order of the processes does not matter. The final line indicates the proportion of CD events excluded over the full dataset.}
    \label{fig:CRC}
    \vspace{-0.3cm}
\end{figure}

Using this pipeline, we extracted 41 candidates from the dataset, which are homogeneously distributed over the studied periods. Their arrival directions are shown in Fig.~\ref{fig:CRC}. 

\begin{figure}[!ht]
    \centering
\includegraphics[width=0.53\textwidth]{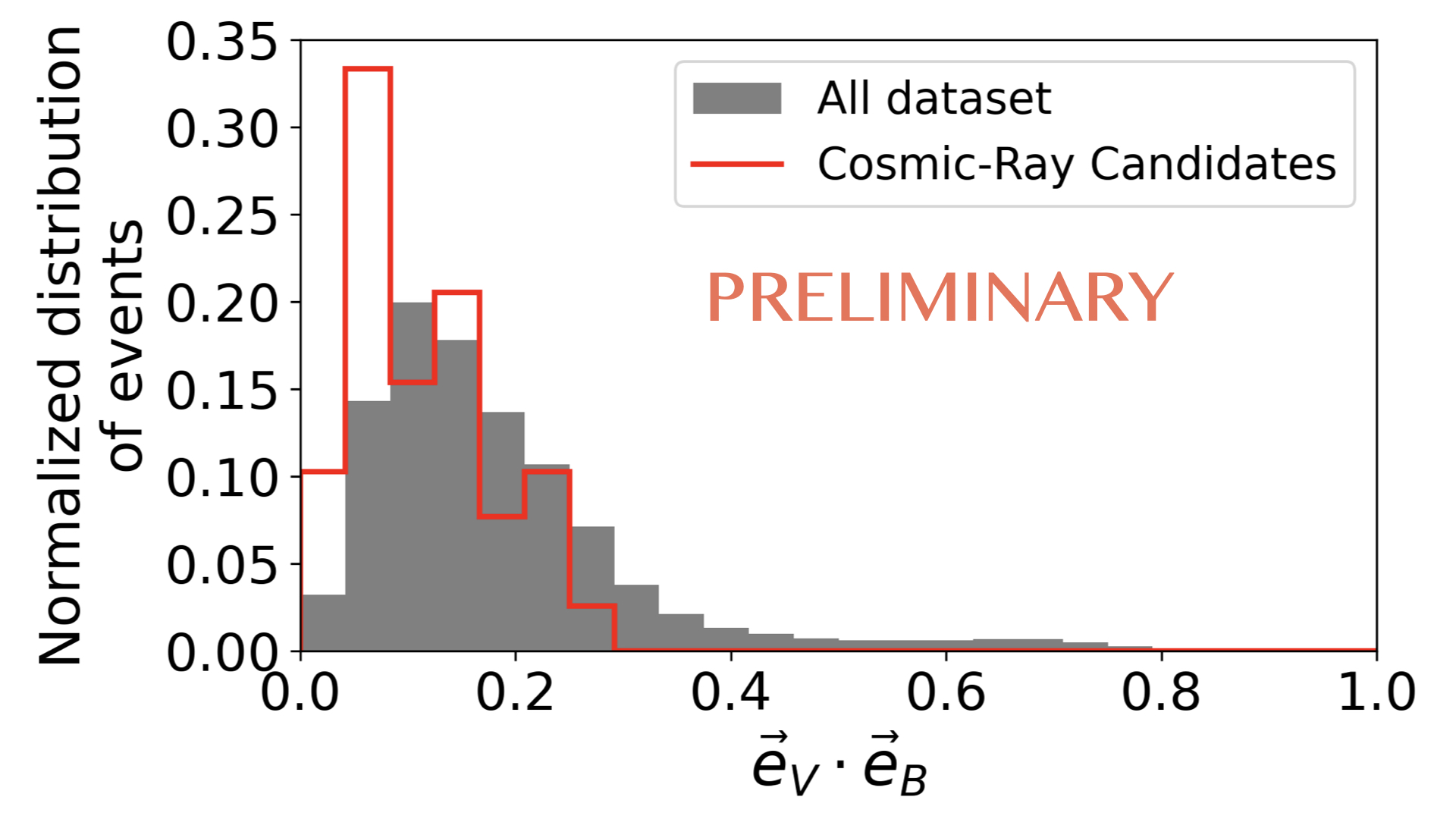}
    \includegraphics[width =0.45\textwidth]{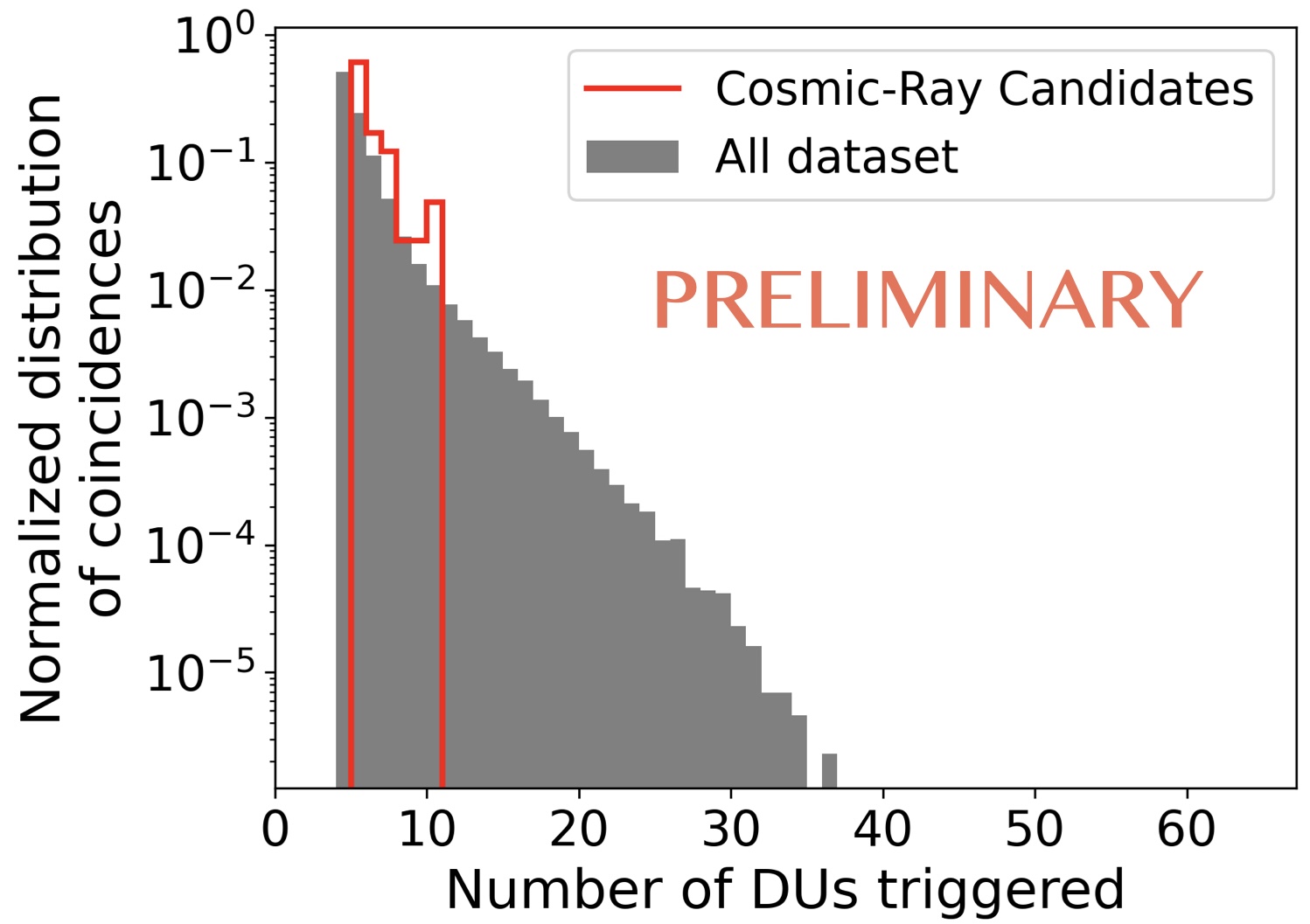}
    \caption{\footnotesize {\bf Left:} Normalized distribution of polarization in cosmic-ray candidates and for all dataset. {\bf Right:} Normalized distribution of the number of antennas triggered for cosmic-ray candidates and all dataset. }
    \label{fig:distrib}
    \vspace{-0.5cm}
\end{figure}

All cosmic-ray candidate events were processed through the reconstruction pipeline, yielding output energies consistent with our exposure calculations, as referenced in ~\cite{Kato:exposure_ICRC25}. Fig.~\ref{fig:distrib} illustrates the polarization distribution (left) and the number of antennas triggered (right). The polarization distribution gives similar results to Fig.~\ref{fig:effvspur} right: the cosmic-ray polarization distribution peaks at lower $\bf{e_V}\cdot\bf{e_B}$ values than the distribution for all dataset. 
The number of antennas triggered serves as an indicator of footprint size, for both the candidates and the overall dataset. The number of triggered antennas aligns with expectations from ~\cite{a2024pruningtooloptimizelayout}, which suggest that the event's footprint should range between $\sim 1-10$ km$^2$ in the cosmic-ray energy and zenith range where GP300 is most sensitive. This corresponds to between 5 and 10 triggered antennas with GP300. After passing these 41 event candidates through reconstruction quality cut (cf. criteria in the previous section), we retrieve 26 solid cosmic-ray candidates, displayed in~\cite{Martineau:GRAND_ICRC25}, Figure 5.

We caution that the distributions of energy or arrival directions should be taken with care and not be interpreted, as this preliminary search was conducted with certain biases: particular attention was given to the Northern part of our dataset when performing manual cuts (from azimuths $\varphi=0-45^\circ$ and $315-360^\circ$), although the other regions were also examined.

\vspace{-0.3cm}
\section{Perspectives}\label{sec3}
\vspace{-0.3cm}
In the near future, GP300 is expected to provide enhanced data with improved trigger efficiency and an increased number of antennas: 65 are already deployed as of July 2025, and once the full GP300 configuration is deployed in 2026, we expect $\sim\,130$ cosmic-ray events per day~\cite{Kato:exposure_ICRC25}. Furthermore, additional physical cuts are planned for implementation, and existing cuts are set to be enhanced. These improvements include the application of machine learning techniques for noise removal and physical feature identification (such as timing or improved polarization cut), utilizing both conventional and machine learning methods.

\bibliographystyle{ICRC}
\setlength{\bibsep}{0pt plus 0.3ex}
{\footnotesize
\bibliography{references}

\providecommand{\href}[2]{#2}\begingroup\raggedright\begin{thebibliography}{10}

\bibitem{Martineau:GRAND_ICRC25}
{\bfseries GRAND} Collaboration, O.~Martineau {\em PoS} {\bfseries ICRC2025} (2025) 1114.

\bibitem{GRANDprotoHW_25}
{\bfseries GRAND} Collaboration, ``Grand prototypes: Grandproto300 and grand@auger,'' 2025.

\bibitem{Zhang-Kewen:electric_recons_ICRC25}
{\bfseries GRAND} Collaboration, K.~Zhang {\em PoS} {\bfseries ICRC2025} (2025) 447.

\bibitem{Kato:exposure_ICRC25}
{\bfseries GRAND} Collaboration, S.~Kato {\em PoS} {\bfseries ICRC2025} (2025) 298.

\bibitem{GRANlib24}
{GRAND Collaboration} \href{http://dx.doi.org/10.1016/j.cpc.2024.109461}{{\em Computer Physics Communications} {\bfseries 308} (Mar., 2025) 109461}.

\bibitem{Correa:NUTRIG_ICRC25}
{\bfseries GRAND} Collaboration, P.~Correa {\em PoS} {\bfseries ICRC2025} (2025) 229.

\bibitem{BenoitLevy:denoising_ICRC25}
{\bfseries GRAND} Collaboration, A.~Benoit-Lévy {\em PoS} {\bfseries ICRC2025} (2025) 185.

\bibitem{Ferri_re_2025}
A.~Ferrière, S.~Prunet, A.~Benoit-Lévy, {\em et~al.} \href{http://dx.doi.org/10.1016/j.nima.2024.170178}{{\em NIM A} {\bfseries 1072} (Mar., 2025) 170178}.

\bibitem{Chiche_2022}
S.~Chiche, K.~Kotera, O.~Martineau-Huynh, M.~Tueros, and K.~D. de~Vries \href{http://dx.doi.org/10.1016/j.astropartphys.2022.102696}{{\em Astropart. Phys.} {\bfseries 139} (2022) 102696}.

\bibitem{Charrier_2019}
D.~Charrier, K.~de~Vries, Q.~Gou, {\em et~al.} \href{http://dx.doi.org/10.1016/j.astropartphys.2019.03.002}{{\em Astroparticle Physics} {\bfseries 110} (July, 2019) 15–29}.

\bibitem{a2024pruningtooloptimizelayout}
A.~Benoit-Lévy, K.~Kotera, and M.~Tueros \href{http://dx.doi.org/10.1088/1748-0221/19/04/P04006}{{\em Journal of Instrumentation} {\bfseries 19} (Apr., 2024) P04006}.

\bibitem{Gulzow:ICRC25}
{\bfseries GRAND} Collaboration, L.~Gülzow {\em PoS} {\bfseries ICRC2025} (2025) 283.

\bibitem{Guelfand:volt_recons_ICRC25}
{\bfseries GRAND} Collaboration, M.~Guelfand {\em PoS} {\bfseries ICRC2025} (2025) 278.

\bibitem{guelfand2025reconstructioninclinedextensiveair}
M.~Guelfand, V.~Decoene, O.~Martineau-Huynh, {\em et~al.}, 2025.
\newblock \url{https://arxiv.org/abs/2504.18257}.

\bibitem{Ferriere:GNN_ICRC25}
{\bfseries GRAND} Collaboration, A.~Ferriere {\em PoS} {\bfseries ICRC2025} (2025) 253.

\end{thebibliography}\endgroup
}

\clearpage

\section*{Full Author List: GRAND Collaboration}

\scriptsize
\noindent
J.~Álvarez-Muñiz$^{1}$, R.~Alves Batista$^{2, 3}$, A.~Benoit-Lévy$^{4}$, T.~Bister$^{5, 6}$, M.~Bohacova$^{7}$, M.~Bustamante$^{8}$, W.~Carvalho$^{9}$, Y.~Chen$^{10, 11}$, L.~Cheng$^{12}$, S.~Chiche$^{13}$
% ...
\\
\\
$^{1}$Departamento de Física de Particulas \& Instituto Galego de Física de Altas Enerxías, Universidad de Santiago de Compostela, 15782 Santiago de Compostela, Spain \\
$^{2}$Institut d'Astrophysique de Paris, CNRS  UMR 7095, Sorbonne Université, 98 bis bd Arago 75014, Paris, France \\
$^{3}$Sorbonne Université, Université Paris Diderot, Sorbonne Paris Cité, CNRS, Laboratoire de Physique  Nucléaire et de Hautes Energies (LPNHE), 4 Place Jussieu, F-75252, Paris Cedex 5, France \\
$^{4}$Université Paris-Saclay, CEA, List,  F-91120 Palaiseau, France \\
$^{5}$Institute for Mathematics, Astrophysics and Particle Physics, Radboud Universiteit, Nijmegen, the Netherlands \\
$^{6}$Nikhef, National Institute for Subatomic Physics, Amsterdam, the Netherlands \\
$^{7}$Institute of Physics of the Czech Academy of Sciences, Na Slovance 1999/2, 182 00 Prague 8, Czechia \\
$^{8}$Niels Bohr International Academy, Niels Bohr Institute, University of Copenhagen, 2100 Copenhagen, Denmark \\
$^{9}$Faculty of Physics, University of Warsaw, Pasteura 5, 02-093 Warsaw, Poland \\
$^{10}$Key Laboratory of Dark Matter and Space Astronomy, Purple Mountain Observatory, Chinese Academy of Sciences, 210023 Nanjing, Jiangsu, China \\
$^{11}$School of Astronomy and Space Science, University of Science and Technology of China, 230026 Hefei Anhui, China \\
$^{12}$National Astronomical Observatories, Chinese Academy of Sciences, Beijing 100101, China \\
$^{13}$Inter-University Institute For High Energies (IIHE), Université libre de Bruxelles (ULB), Boulevard du Triomphe 2, 1050 Brussels, Belgium \\
% ...


\subsection*{Acknowledgments}

\noindent
The GRAND Collaboration is grateful to the local government of Dunhuag during site survey and deployment approval, to Tang Yu for his help on-site at the GRANDProto300 site, and to the Pierre Auger Collaboration, in particular, to the staff in Malarg\"ue, for the warm welcome and continuing support.
The GRAND Collaboration acknowledges the support from the following funding agencies and grants.
\textbf{Brazil}: Conselho Nacional de Desenvolvimento Cienti\'ifico e Tecnol\'ogico (CNPq); Funda\c{c}ão de Amparo \`a Pesquisa do Estado de Rio de Janeiro (FAPERJ); Coordena\c{c}ão Aperfei\c{c}oamento de Pessoal de N\'ivel Superior (CAPES).
\textbf{China}: National Natural Science Foundation (grant no.~12273114); NAOC, National SKA Program of China (grant no.~2020SKA0110200); Project for Young Scientists in Basic Research of Chinese Academy of Sciences (no.~YSBR-061); Program for Innovative Talents and Entrepreneurs in Jiangsu, and High-end Foreign Expert Introduction Program in China (no.~G2023061006L); China Scholarship Council (no.~202306010363); and special funding from Purple Mountain Observatory.
\textbf{Denmark}: Villum Fonden (project no.~29388).
\textbf{France}: ``Emergences'' Programme of Sorbonne Universit\'e; France-China Particle Physics Laboratory; Programme National des Hautes Energies of INSU; for IAP---Agence Nationale de la Recherche (``APACHE'' ANR-16-CE31-0001, ``NUTRIG'' ANR-21-CE31-0025, ANR-23-CPJ1-0103-01), CNRS Programme IEA Argentine (``ASTRONU'', 303475), CNRS Programme Blanc MITI (``GRAND'' 2023.1 268448), CNRS Programme AMORCE (``GRAND'' 258540); Fulbright-France Programme; IAP+LPNHE---Programme National des Hautes Energies of CNRS/INSU with INP and IN2P3, co-funded by CEA and CNES; IAP+LPNHE+KIT---NuTRIG project, Agence Nationale de la Recherche (ANR-21-CE31-0025); IAP+VUB: PHC TOURNESOL programme 48705Z. 
\textbf{Germany}: NuTRIG project, Deutsche Forschungsgemeinschaft (DFG, Projektnummer 490843803); Helmholtz—OCPC Postdoc-Program.
\textbf{Poland}: Polish National Agency for Academic Exchange within Polish Returns Program no.~PPN/PPO/2020/1/00024/U/00001,174; National Science Centre Poland for NCN OPUS grant no.~2022/45/B/ST2/0288.
\textbf{USA}: U.S. National Science Foundation under Grant No.~2418730.
Computer simulations were performed using computing resources at the CCIN2P3 Computing Centre (Lyon/Villeurbanne, France), partnership between CNRS/IN2P3 and CEA/DSM/Irfu, and computing resources supported by the Chinese Academy of Sciences.

\end{document}